# Light-induced metastability in $Cu(In,Ga)Se_2$ caused by $V_{Se}$-$V_{Cu}$ complexes

Stephan Lany and Alex Zunger

*National Renewable Energy Laboratory, Golden, CO 80401*

## ABSTRACT

Using first-principles total-energy calculations we identify the microscopic origin and the physical mechanism that leads to light-induced metastability and persistent photoconductivity in the photovoltaic material $Cu(In,Ga)Se_2$. In the presence of photoexcited or electrically injected conduction band electrons, the complex ($V_{Se}$-$V_{Cu}$) of a Se vacancy with a Cu vacancy is predicted to transform from a shallow electron trap into a deep hole trap upon the persistent capture of two electrons. This explains the experimental evidence that for every two holes persistently released to the valence band, one hole trap is generated.

While extensively studied during the past decades [T1, 2, 3, 4], light-induced metastability is a widely observed, yet still puzzling phenomenon in $Cu(In,Ga)Se_2$ (CIGS) based solar cell devices. Whereas metastability is often detrimental to other devices (e.g., the Staebler-Wronski degradation in *a*-Si [5]), in CIGS the light-induced transition into the metastable state has a beneficial effect on the performance of the devices [6], as light-soaking re-establishes the initial state after the conductivity has deteriorated due to thermal treatment [2]. The metastability manifests itself by relaxation of the open circuit voltage and changes in the junction capacitance, which can be explained as persistent photoconductivity (PPC) [7]. Similar metastable changes occur also after minority carrier (electron) injection due to pulses of forward bias [2, 8], where the equivalence of the effect illumination and electron injection indicates that in both cases metastability is caused by the same mechanism. The ubiquitous occurrence of these effects in both single-crystal and in thin-film devices, as well as in films without heterojunction [2] provides strong indication that the phenomenon of metastability in CIGS is an intrinsic property of the bulk material. However, despite the progress in the characterization of the light-induced effects, the microscopic origin of metastability and the physical mechanism that leads to persistent carrier capture and PPC after both illumination and electron injection is still obscure [4, 9].

Using first-principles electronic structure calculations, we have recently shown [10] that the anion vacancy $V_{Se}$ in $CuInSe_2$ and $CuGaSe_2$ can bind two electrons in a deep level, which originates from the formation of a metal-metal (In-In, Ga-Ga) bond between the cation neighbors of $V_{Se}$. This level can cause persistent photoconductivity via the metastable increase of the concentration of free holes when electrons from the valence band maximum (VBM) are transferred to the deep level [10]. Here, we focus on the ($V_{Se}$-$V_{Cu}$) *complex* formed by the Se vacancy with a Cu vacancy: Our calculated configuration coordinate diagram predicts that the ($V_{Se}$-$V_{Cu}$) complex converts, upon illumination, from being a shallow donor (i.e. an electron trap) into a moderately deep acceptor (i.e. a hole trap) by the persistent capture of two electrons (i.e. release of two holes). This reconciles the experimental observation that PPC is accompanied by the appearance of a deep hole trap, where the increase of free carrier and deep acceptor concentration appears in a 1:1 ratio [8, 11].

We determine the defect formation energies $\Delta H_{D,q}$ (D = $V_{Se}$, $V_{Cu}$, $V_{Se}$-$V_{Cu}$) from total energy supercell (64 atoms) calculations in the pseudopotential-momentum space formalism [12] using the projector augmented wave (PAW) potentials [13] and the local density approximation. Using the calculated total energies, we obtain $\Delta H_{D,q}$ as

$$\Delta H_{\mathrm{D},q}(E_\mathrm{F},\mu) = (E_{\mathrm{D},q} - E_\mathrm{H}) + \sum_\alpha \mu_\alpha + qE_\mathrm{F}, \qquad (1)$$

where the first term is the energy difference between the supercell with defect in charge state $q$ ($E_{D,q}$) and the pure host ($E_H$). The second term describes the chemical reservoir of the atoms $\alpha$ ($\alpha$ = Se, Cu) removed from the lattice to form the vacancies, where the chemical potentials $\mu_\alpha = \mu_\alpha^{\mathrm{elem}} + \Delta\mu_\alpha$ are given with respect to the elemental phase. The third term in Eq. (1) is the energy of the electron reservoir, i.e. the Fermi energy $E_\mathrm{F} = E_\mathrm{v} + \Delta E_\mathrm{F}$ given with respect to the energy $E_v$ of the VBM. From the calculated defect formation energies, we also obtain the defect transition energy $\varepsilon(q/q')$ between two charge states $q$ and $q'$, i.e. the value of the Fermi energy where $\Delta H(q, E_F) = \Delta H(q', E_F)$. We apply a recently developed scheme [14] of total-energy corrections for defect calculations to cope with the underestimated LDA band gap and finite supercell size effects such as the spurious interaction of periodic image charges and the Moss-Burstein like band-filling that occurs when electrons or holes occupy a host-like impurity band.

*Electronic states of the isolated Se vacancy*. The neutral vacancy $V_{Se}^0$ introduces two electronic levels in the band-gap region, as shown schematically in Fig. 1a: The state denoted *a* results from the *bonding* state formed by the two In-site centered dangling bonds, and is located below the VBM, while the state denoted *b* results from the *antibonding* state formed by the two dangling bonds, and is located within the band gap. The neutral $V_{Se}^0$ has two electrons, but the levels *a* and *b* can be occupied by up to 4 electrons, which means that the possible charge states range from 2+ to 2–, and that $V_{Se}$ can be amphoteric. Our total energy calculations show that only the even charge states are thermodynamically stable due to "negative-*U*" behavior, i.e. the odd charge states disproportionate exothermically according to $2\,V_{Se}^- \rightarrow V_{Se}^0 + V_{Se}^{2-}$, and $2\,V_{Se}^+ \rightarrow V_{Se}^0 + V_{Se}^{2+}$.

The calculated ε(2+/0) donor and ε(0/2–) acceptor transition levels are given in Table I for CIS and CGS. Both transitions have important implications for CIGS based solar cells: First, the thermal ε(2+/0) transition ($a^0 \rightarrow a^2$) is accompanied by an activated change in the atomic structure of $V_{Se}$, and leads to metastability and PPC. This is discussed in detail below. Second, the *b* level associated with the thermal ε(0/2–) transition ($a^2b^0 \rightarrow a^2b^2$) can act as a recombination center and possibly affects the solar cell performance. We calculated the optical absorption energy $E_{abs}$ corresponding to the excitation of an electron from the VBM into the *b* level ($a^2b^0 \rightarrow a^2b^1 + h$). We see in Table I that this absorption energy is close to the band gap energy in CIS ($E_g$ = 1.04 eV), and is only slightly higher in CGS despite the larger band gap ($E_g$ = 1.68 eV). The energy and the fact that the this absorption level is almost constant with respect to the VBM when comparing CIS and CGS resembles the properties of an experimentally observed absorption level around 0.8 eV which has been discussed as an important recombination center in CIGS solar cells [15]. Since we predict that the metastable behavior arises from the activated change between the $V_{Se}^0$ and $V_{Se}^{2+}$ charge states (see discussion below), and that the absorbing level *b* is present in the gap only in the $V_{Se}^0$ state (Fig. 1a), it should be possible to determine experimentally whether or not the observed level is related to $V_{Se}$ by testing the correlation with the metastable behavior of CIGS.

*Metastability of $V_{Se}$*. The neutral Se vacancy in CIS and CGS has short In-In (Ga-Ga) metal-metal bonds with $d_{\text{In-In}}$ = 3.04 Å ($d_{\text{Ga-Ga}}$ = 2.83 Å) whereas $V_{Se}^{2+}$ exhibits a breakup of these bonds with distant cations, $d_{\text{In-In}}$ = 5.45 Å ($d_{\text{Ga-Ga}}$ = 5.29 Å). This structural change is accompanied by a dramatic change of the electronic energy levels, as shown schematically in Fig. 1a. Both the *a* and *b* levels move into the conduction band. In order to discuss the detailed physical mechanism of metastability and PPC, arising from this behavior, we show in Fig. 1b the calculated configuration coordinate diagram, i.e. the formation energies Δ*H* as a function of the In-In bond length for the (charge conserving) vacancies $V_{Se}^0 + 2h$, $V_{Se}^+ + h$, and $V_{Se}^{2+}$. The dashed lines correspond to LDA energies which yield underestimated Δ*H* when the occupied *a* level is above the calculated CBM. The solid lines give the Δ*H* curves for corrected [14] energies. Since the initially empty ($V_{Se}^{2+}$) or partially occupied ($V_{Se}^+ + h$) *a* level shifts to energies below the VBM for $d_{\text{In-In}} < 3.9$ Å (cp. Fig. 1a), it becomes occupied by valence band electrons. Thus, $V_{Se}^{0}$ + 2*h* is formed in both cases, and the branches of the different charge states in Fig. 1b coincide for close In-In distances. With increasing In-In distance, the shift of the *a* level higher energies leads to high formation energies in the neutral state $V_{Se}^{0}$, where the *a* level is always occupied. In the ionized state $V_{Se}^{2+}$, the *a* level becomes unoccupied when it crosses the VBM ($V_{Se}^{0}$ + 2*h* → $V_{Se}^{2+}$), and the system relaxes to a new equilibrium structure with large In-In distance of 5.45 Å. This leads to the double-well structure of the $V_{Se}^{2+}$ branch in Fig. 1b. The energy of the $V_{Se}^{+}$ state rises with $d_{In\text{-}In}$ at a lower rate than that of $V_{Se}^{0}$ because the *a* level is only singly occupied. Since this level crosses the (corrected) CBM at $d_{In\text{-}In} \approx 4.8$ Å, the electron can relax to the CBM ($V_{Se}^{+}$ + *h* → $V_{Se}^{2+}$ + *e* + *h*), leading again to a double well.

Having constructed the configuration coordinate diagram, we now explain the mechanism leading to PPC: First, above-gap illumination (excitation "opt." in Fig. 1) creates free electron-hole pairs. Subsequently, the doubly charged vacancy $V_{Se}^{2+}$ can change into the singly charged state $V_{Se}^{+}$ by thermal activation across the energy barrier $\Delta E_1 \approx 0.1$ eV in Fig. 1b and simultaneous capture of a photoexcited electron ($V_{Se}^{2+}$ + *e* + *h* → $V_{Se}^{+}$ + *h*). Atomic relaxation leads, without further barrier, to the short In-In distance of 3.04 Å (Fig. 1b), where the *a* level is below the VBM and captures another electron from the VBM ($V_{Se}^{+}$ + *h* → $V_{Se}^{0}$ + 2*h*). Thus, illumination initiates the overall reaction $V_{Se}^{2+}$ → $V_{Se}^{0}$ + 2*h*, where the two released holes increase the *p*-type conductivity. This constitutes PPC. Thermal activation across the second barrier $\Delta E_2 = 0.35$ eV and simultaneous capture of two holes leads to the back-transition to $V_{Se}^{2+}$ and the loss of PPC. Despite the moderate barrier height, this process occurs rather slowly, because of the requirement that two holes must be captured. For $V_{Se}$ in CGS, we find a very similar energy diagram with one significant difference: Due to the larger band gap, the *a* level does not shift to energies significantly above the CBM in the doubly charged state. This leads to a vanishing barrier $\Delta E_1$. Therefore, while it might be possible in CIS to freeze-out the light-induced effect at low temperature due to the first barrier $\Delta E_1$, this may not be possible in CGS. The barrier $\Delta E_2$ exists like in CIS, but is somewhat smaller, i.e. $\Delta E_2 = 0.25$ eV.

Our model of PPC caused by the isolated Se vacancy explains two important experimental observations: First, since it is only required that $V_{Se}^{2+}$ captures an electron from the conduction band, illumination and electron injection have similar effect [2, 8]. Second, the photon energy threshold for PPC practically equals the band gap energy. During scanning the photon energy, the dominant response of PPC was found for above band-gap energies [9]. A sub-gap response around 0.7 eV with much smaller intensity was also observed in Ref. [9]. This is probably due to excitation of electrons from deep acceptor states to the conduction band, which requires less energy than the band gap. The model of the isolated Se vacancy cannot explain, however, the occurrence of a hole trap associated with the electron capture [4, 8]. This requires consideration of the vacancy complex.

*Formation of ($V_{Se}$-$V_{Cu}$) vacancy complexes*. Figure 2a shows the calculated binding energy $E_b = \Delta H(V_{Se}) + \Delta H(V_{Cu}) - \Delta H(V_{Se}\text{-}V_{Cu})$ of the complex, which is, e.g., $E_b = 0.4$ eV at $\Delta E_F = 0.2$ eV in CIS ($E_b = 0.6$ eV at $\Delta E_F = 0.2$ eV in CGS). Thus, $V_{Se}$ binds $V_{Cu}$. Since copper is known to be mobile in CIS even at room temperature [6, 16], we expect that $V_{Cu}$ will equilibrate to form the $V_{Se}$-$V_{Cu}$ complex at room temperature. From the law of mass action, i.e. $[(V_{Se}\text{-}V_{Cu})] = [V_{Se}]\times[V_{Cu}]\times\exp(E_b/kT)$, we expect that practically all Se vacancies will be present as complexes if the $V_{Cu}$ concentration exceeds $10^{16}$cm$^{-3}$, which is generally the case in these materials.

Figure 2b shows as a function of the Fermi level the calculated defect formation energies and transition energies (dots) for $V_{Se}$, $V_{Cu}$, and the ($V_{Se}$-$V_{Cu}$) vacancy complex in CIS for Se-poor ($\Delta\mu_{Se}$ = –0.83 eV [17]) and Cu-rich ($\Delta\mu_{Cu}$ = 0) conditions. While 0 and 2+ are the stable charge states for $V_{Se}$, the stable states of the complex are shifted to 1– and 1+, owing to the acceptor property of $V_{Cu}$. The ε(+/–) transition energy ($a^0 \rightarrow a^2$) is around $E_v$ + 0.3 eV (Table I and Fig. 2), so that the 1+ state is the equilibrium stable state for Fermi levels typically present at room temperature in *p*-type material.

The mechanism of III-III bond formation [for $\left(V_{Se}\text{-}V_{Cu}\right)^-$] and breakup [for $\left(V_{Se}\text{-}V_{Cu}\right)^+$] is still operational in the complex, and the electronic structure is only little influenced by the presence of $V_{Cu}$: the respective *a* and *b* levels, schematically shown in Fig. 3a, are shifted only slightly to higher energies. This means that the optical absorption level and metastability occur in a similar fashion as for the isolated $V_{Se}$. PPC results from the light-induced transition $\left(V_{Se}\text{-}V_{Cu}\right)^+ \rightarrow \left(V_{Se}\text{-}V_{Cu}\right)^- + 2h$. From knowledge of the configuration coordinate diagram of the isolated $V_{Se}$ and the calculated total energies of the equilibrium structures of the complex in the different charge states, we construct in Fig. 3b the configuration coordinate diagram of the vacancy complex. The important new feature relative to the isolated $V_{Se}$ is that the negatively charged $\left(V_{Se}\text{-}V_{Cu}\right)^-$ complex now can attract and bind a hole, forming an hole trap. This hole trap is present only in the light-induced metastable configuration of the vacancy complex, and corresponds to the ε(0/–) acceptor level of this configuration. The calculated hole trap depth $E_a$ = $E_v$ + 0.27 eV in CIS (Table I and Fig. 3) is in excellent agreement with experimental data [8]. Notably, this level is shallower in CGS (cp. Table I), having a transition energy more similar to that of the isolated $V_{Cu}$. The presence of the *b* level (Fig. 3a) in the gap could increase carrier recombination in the metastable state. The overall beneficial effect of light-soaking [2, 6] indicates, however, that the effect of increased carrier concentration due to PPC outweighs the effect of increased recombination.

We summarize the mechanism of metastability in CIGS as follows: Before illumination, the positively charged vacancy complex $\left(V_{Se}\text{-}V_{Cu}\right)^+$ is the thermodynamically stable state in *p*-type material. Due to the positive charge, this center acts as a shallow *electron trap*. After capturing a photoexcited or injected electron (which requires thermal activation across the barrier $\Delta E_1$ in Fig. 3b), the complex undergoes a structural transformation leading to formation of short In-In (Ga-Ga) bonds. During the transformation into this metastable state, a second electron is captured from the valence band, so that in total two free holes are released, which leads to PPC. At the same time, the complex becomes negatively charged. This $\left(V_{Se}\text{-}V_{Cu}\right)^-$ state acts as a *hole trap*. The PPC decays in the course of hole capture due to thermal activation across the barrier $\Delta E_2$ in Fig. 3b, leading to the back-transition into the positively charged ground state of the complex.

This work was supported by DOE-EERE, under Grant No. DEAC36-98-GO10337.

Table I: Calculated transition energies in eV for the isolated $V_{Se}$ and the ($V_{Se}$-$V_{Cu}$) complex in CIS and CGS. For $V_{Se}$, we give the thermal ε(2+/0) double donor and ε(0/2–) double acceptor levels, and the optical absorption level $E_{abs}$ of $V_{Se}^{0}$. For the complex, we give the ε(+/–) transition energy between the equilibrium stable 1+ and 1– states, and the hole trap energy $E_a$, i.e. the ε(0/–) acceptor level of the light-induced metastable configuration.

| | Isolated $V_{Se}$ | | | ($V_{Se}$-$V_{Cu}$) | |
|---|---|---|---|---|---|
| | ε(2+/0) | ε(0/2–) | $E_{abs}$ | ε(+/–) | $E_a$ |
| CIS | $E_v$+ 0.02 | $E_v$+ 0.97 | $E_v$+ 1.02 | $E_v$+ 0.31 | $E_v$+ 0.27 |
| CGS | $E_v$+ 0.08 | $E_v$+ 1.10 | $E_v$+ 0.98 | $E_v$+ 0.28 | $E_v$+ 0.03 |

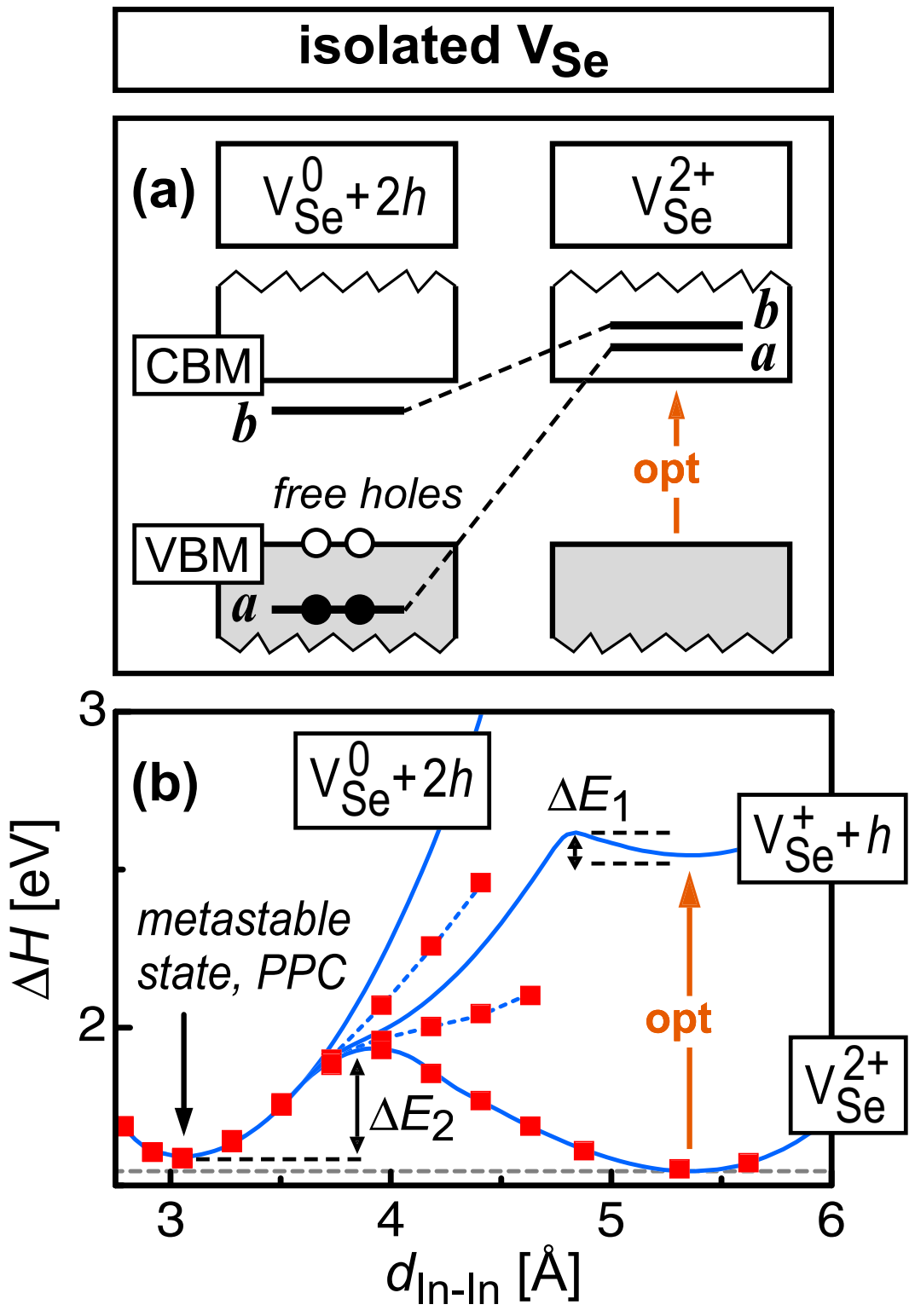

Figure 1 (color online): Energy level (a) and configuration coordinate (b) diagrams for the isolated Se vacancy in CIS. The calculated defect formation energies Δ*H* (red squares) refer to Se-poor conditions and $E_F = E_v$.

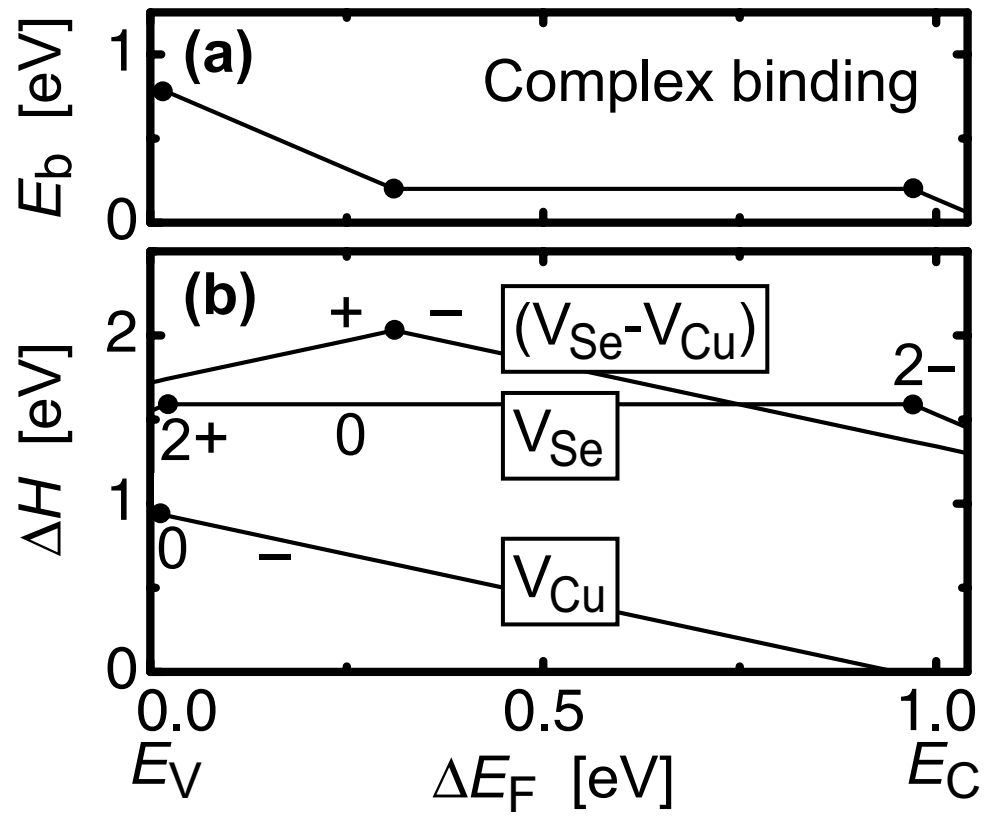

Figure 2. (a) Binding energy of the ($V_{Se}$-$V_{Cu}$) complex. (b) Defect formation energies of $V_{Se}$, $V_{Cu}$ and ($V_{Se}$-$V_{Cu}$) in CIS under Se-poor and Cu-rich growth conditions, as a function of $\Delta E_F$.

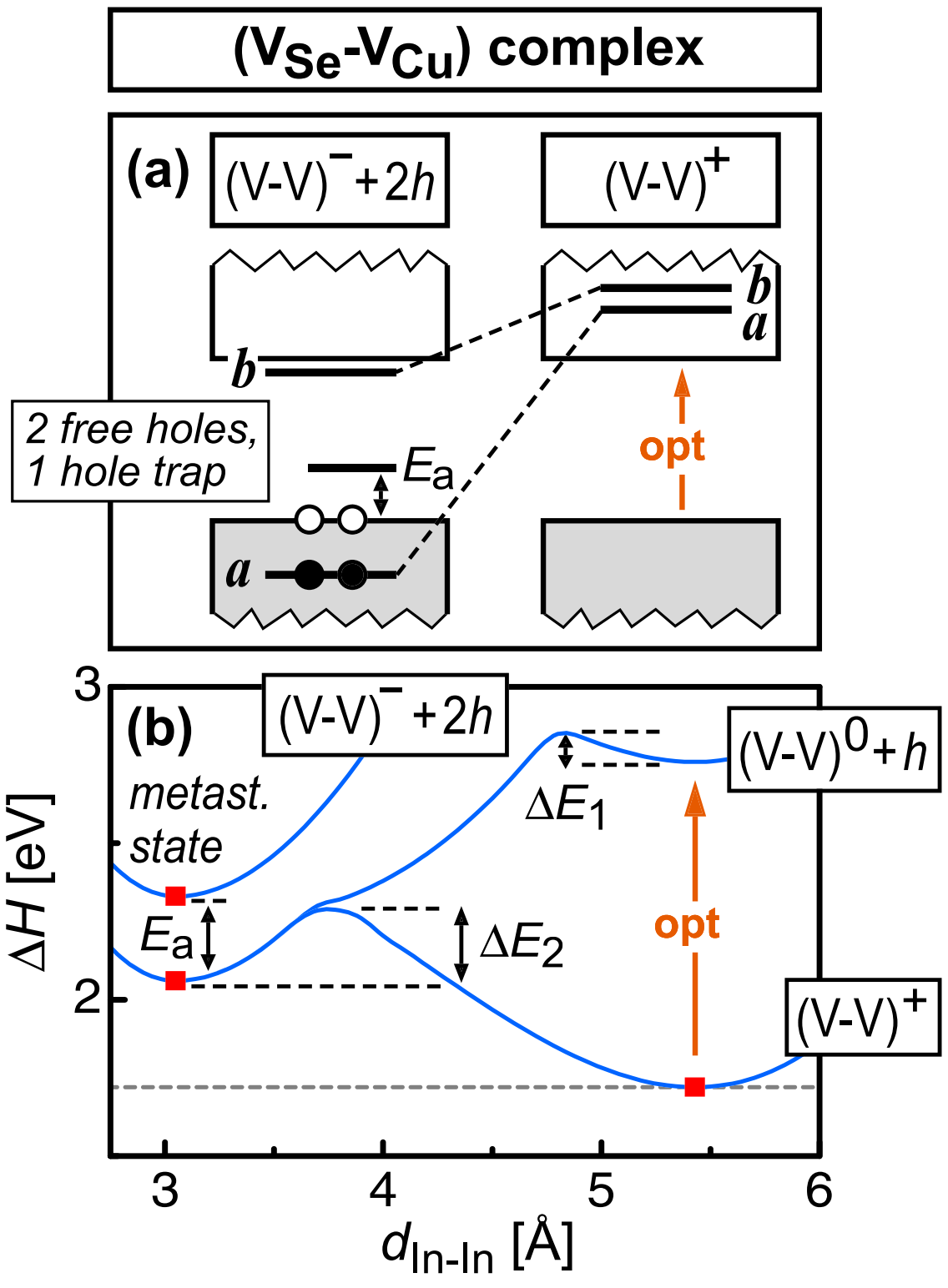

Figure 3: (color online): Energy level (a) and configuration coordinate (b) diagrams for the ($V_{Se}$-$V_{Cu}$) vacancy complex in CIS. $E_a$ indicates the hole trap energy in the metastable configuration. The calculated defect formation energies $\Delta H$ (red squares) refer to Se-poor and Cu-rich conditions and $E_F = E_v$.